\documentclass[prd,aps,floats,twocolumn,showpacs]{revtex4}
\usepackage[dvips]{epsfig}

\begin{document}

\def\sqr#1#2{{\vcenter{\hrule height.#2pt
   \hbox{\vrule width.#2pt height#1pt \kern#1pt
      \vrule width.#2pt}
   \hrule height.#2pt}}}
\def\square{{\mathchoice\sqr64\sqr64\sqr{3.0}3\sqr{3.0}3}}

\title{Extended gauge invariance and electroweak interactions}

\author{Bernd A. Berg}

\affiliation{Department of Physics, Florida State
             University, Tallahassee, FL 32306-4350, USA}

\date{November 1, 2009.} % \date{\today } 

\begin{abstract}
Gauge invariance is extended to allow to allow for a  U(1)~--~SU(2)
mixing term, which can cause a SU(2) deconfining transition.
\end{abstract}
\pacs{12.15.-y, 12.60.Cn, 12.60.-i, 14.80.Bn, 11.15.Ha}
\maketitle

Recently, it was shown in a lattice gauge theory simulation \cite{Be09}
that a U(1)~--~SU(2) interaction term can cause a SU(2) deconfining 
phase transition quite similar to the confinement-Higgs transition 
ovserved in \cite{Early}. This interaction requires unusual gauge 
transformations, which are written down here for the continuum
formulation. So, the subsequent construction is motivated by
making the interaction
\begin{eqnarray} \label{Lint} 
  L^{\rm int} &=& -\frac{\lambda}{2}{\rm Tr}\left(
  F^{\rm int}_{\mu\nu}F^{\rm int}_{\mu\nu}\right)\,,\\ \label{Fint}
  F^{\rm int}_{\mu\nu} &=& g_a\partial_{\mu}A_{\nu} -
  g_b\partial_{\nu}B_{\mu} + i\,g_ag_b\left[B_{\mu},A_{\nu}\right]\,.
\end{eqnarray}
gauge invariant (the commutator term is missing in \cite{Be09}, because 
the continuum limit was extracted there in a gauge in which $A_{\mu}$ 
is diagonal). The interaction (\ref{Lint}) leads to a mixing of U(1) 
and SU(2) fields as found in the standard model after the Higgs 
mechanism. However, it is presently not clear whether our approach 
will altogether lead to similar physics.

In the following we use Euclidean notation. Let us first consider 
the usual gauge-covariant derivative 
\begin{equation} \label{Dem} 
  D^{\rm em}_{\mu} = \partial_{\mu} + ig a_{\mu}
\end{equation}
of an electromagnetic field $a_{\mu}(x)$ on a complex fermion field 
$\psi(x)$. With the gauge transformations
\begin{eqnarray} 
  \psi &\to& \psi' = e^{i\alpha(x)}\,\psi\,, \\
  a_{\mu}(x)&\to& a'_{\mu}=a_{\mu}-\frac{1}{g}\,\partial_{\mu}\alpha(x)
\end{eqnarray}
one finds
\begin{equation}
  D^{\rm em}_{\mu}\psi\to D^{'\rm em}_{\mu}\psi'
  = e^{i\alpha(x)}D^{\rm em}_{\mu}\psi\,,
\end{equation}
so that the Lagrangian
\begin{equation} \label{Lem} 
  L^{\rm em} = \overline{\psi}\left(i\gamma_{\mu}D^{em}_{\mu}-m\right)
               \psi -\frac{1}{4}F^{\rm em}_{\mu\nu}F^{\rm em}_{\mu\nu}
\end{equation}
is gauge invariant, where 
\begin{equation} \label{Fem} 
  F^{\rm em}_{\mu\nu}=\frac{1}{ig}
  \left[D^{em}_{\mu},D^{em}_{\nu}\right]\,.
\end{equation}
is the field tensor.

Assume now that $\psi(x)$ is a complex doublet, which transforms
under U(1)$\otimes$SU(2) gauge transformations
\begin{eqnarray} \label{psig}
  \psi\ &\to& \psi' = G(x)\,\psi\,,\\ \label{G}
  G(x) &=& \exp\left(\frac{i}{2}
  \sum_{i=0}^3\tau_i\alpha_i(x)\right)\,.
\end{eqnarray}
Here $\tau_0$ is the $2\times 2$ unit matrix and $\tau_i$, $i=1,2,3$
are the Pauli matrices. We still can couple $\psi$ in a gauge invariant 
way to an electromagnetic field. We define the gauge covariant 
derivative by
\begin{equation} \label{Da} 
  D^a_{\mu} = \partial_{\mu} + ig_a A_{\mu}~~{\rm with}~~
  A_{\mu} = \frac{1}{2}\,\tau_0\,a_{\mu}
\end{equation}
and gauge transformation of $A_{\mu}$ by (compare, e.g., \cite{Qu83})
\begin{equation} \label{Ag} 
  A_{\mu}\to A'_{\mu} = GA_{\mu}G^{-1}
  + \frac{i}{g_a}(\partial_{\mu} G)G^{-1}
\end{equation}
which adds the transformation of a null SU(2) field and yields the 
desired result
\begin{equation} \label{Dpsig}
  D^a_{\mu}\psi\to D^{'a}_{\mu}\psi' = G D^a_{\mu}\psi\,.
\end{equation}
The field tensor defined by
\begin{equation} \label{Fa} 
  F^a_{\mu\nu}=\frac{1}{ig_a}\left[D^a_{\mu},D^a_{\nu}\right]
\end{equation}
transforms as
\begin{equation} \label{Fag} 
  F^a_{\mu\nu}\to F^{'a}_{\mu\nu} = G F^a_{\mu\nu} G^{-1}
\end{equation}
so that the Lagrangian 
\begin{equation} \label{La} 
  L^a = \overline{\psi}\left(i\gamma_{\mu}D^a_{\mu}-m\right)\psi 
      -\frac{1}{2}{\rm Tr}\left(F^a_{\mu\nu}F^a_{\mu\nu}\right)
\end{equation}
stays gauge invariant. The U(1)$\otimes$SU(2) gauge transfromations do
not destroy the fact that $A_{\mu}$ describes just an electromagnetic 
field. Any $A_{\mu}(x)$ field is gauge equivalent to one for which
$A_{\mu}$ is diagonal. Consequently, $F^{'a}_{\mu\nu}$ stays always 
diagonal and the $G$ matrices could be omitted in (\ref{Fag}). Note 
that all arguments hold as well for U(1)$\otimes$SU(N).

Similarly, we can extend gauge transformations of a SU(2) field
$B_{\mu}(x)$ by a phase and couple it with the complex doublet 
(\ref{psig}). The gauge covariant derivative is
\begin{equation} \label{Db} 
  D^b_{\mu} = \partial_{\mu} + ig_b B_{\mu}~~{\rm with}~~
  B_{\mu} = \frac{1}{2}\,\vec{\tau}\cdot\vec{b}_{\mu}
\end{equation}
and the gauge transformation of $B_{\mu}$ are
\begin{equation} \label{Bg} 
  B_{\mu}\to B'_{\mu} = GB_{\mu}G^{-1}
  + \frac{i}{g_b}(\partial_{\mu} G)G^{-1}\,.
\end{equation}
Equations (\ref{Dpsig}) to (\ref{La}) carry simply over by
replacing all labels $a$ by~$b$.

An electroweak Lagrangian of the type
\begin{eqnarray} \label{Lew}
  L &=& -\frac{1}{2}{\rm Tr}\left(F^a_{\mu\nu}F^a_{\mu\nu}
  \right) - \frac{1}{2}\left(F^b_{\mu\nu}F^b_{\mu\nu}\right)
  \\ \nonumber
  &+& \overline{\psi}\left(i\gamma_{\mu}D^a_{\mu}-m\right)\psi 
   +  \overline{\psi}\left(i\gamma_{\mu}D^b_{\mu}-m\right)\psi
\end{eqnarray}
allows then to add the interaction term (\ref{Lint}), for which 
a zero-temperature SU(2) deconfining phase transition was found 
in \cite{Be09}. Under gauge transformations (\ref{Ag}) for $A_{\mu}$ 
and (\ref{Bg}) for $B_{\mu}$, the $F^{\rm int}_{\mu\nu}$ tensor 
(\ref{Fint}) transforms according to 
\begin{equation} \label{Fintg} 
  F^{\rm int}_{\mu\nu}\to F^{'\rm int}_{\mu\nu} = 
  G\, F^{\rm int}_{\mu\nu}\, G^{-1}\,,
\end{equation}
so that $L^{\rm int}$ is gauge invariant. The algebra for (\ref{Fintg}) 
is given in the appendix.

The interaction ({\ref{Lint}) is expected to result in a mixing of 
$a_{\mu}$ and $b^3_{\mu}$ into orthogonal combinations
\begin{eqnarray} \label{gammaA} 
  A^{\gamma}_{\mu} &=& +a_{\mu}\cos\phi + b^3_{\mu}\sin\phi\,,\\
  \label{Z} Z_{\mu}&=& -a_{\mu}\sin\phi + b^3_{\mu}\cos\phi\,,
\end{eqnarray}
where estimating the mixing angle may require non-perturbative
methods due to the nature of the SU(2) deconfining phase 
transition. 
% Note that $\tau_0$ in (\ref{Da}) becomes in this 
% context the hypercharge operator $Y$.

\acknowledgments 
This research was in part supported by the DOE grant DE-FG02-97ER41022 
and by a Humboldt Research Award. Some of the work was done at Leipzig 
University and I am indebted to Wolfhard Janke and his group for their 
kind hospitality. % \bigskip

\appendix \section{Gauge Invariance of $L^{\rm int}$.}

Extending the calculation of \cite{Qu83} slightly, we find
\begin{eqnarray} 
  &~& g_a\partial_{\mu}A'_{\nu} - g_b\partial_{\nu}B'_{\mu} = 
  \\ \nonumber
  &~& \partial_{\mu}[Gg_aA_{\nu}G^{-1}+i(\partial_{\nu}G)G^{-1}] 
  \\ \nonumber 
  &-& \partial_{\nu}[Gg_bB_{\mu}G^{-1}+i(\partial_{\mu}G)G^{-1}]
  =\\ \nonumber &~& \\ 
  &~& G(g_a\partial_{\mu}A_{\nu} - g_b\partial_{\nu}B_{\mu})G^{-1}
  \\ \nonumber 
  &+& [(\partial_{\mu}G)g_aA_{\nu}-(\partial_{\nu}G)g_bB_{\mu}]G^{-1}
  \\ \nonumber &+& 
  G[g_aA_{\nu}(\partial_{\mu}G^{-1})-g_bB_{\mu}(\partial_{\nu}G^{-1})]
  \\ \nonumber &+& 
  i\,[(\partial_{\nu}G)(\partial_{\mu}G^{-1})-
      (\partial_{\mu}G)(\partial_{\nu}G^{-1})]
\end{eqnarray}
Using $(\partial_{\mu}G^{-1})G+G^{-1}(\partial_{\mu}G)=\partial_{\mu}
(G^{-1}G)=0$, this can be transformed to
\begin{eqnarray} \label{dApdBp}
  &~&g_a\partial_{\mu}A'_{\nu} - g_b\partial_{\nu}B'_{\mu} =\\ \nonumber
  &~& G\left(g_a\partial_{\mu}A_{\nu} - g_b\partial_{\nu}B_{\mu}
       \right)G^{-1} \\ \nonumber &+&
  G\left\{\left[G^{-1}(\partial_{\mu}G),g_aA_{\nu}\right] -
  \left[G^{-1}(\partial_{\nu}G),g_bB_{\mu}\right]\right\}G^{-1}
  \\ \nonumber 
  &-& i\,G[(\partial_{\mu}G^{-1})(\partial_{\nu}G) -
           (\partial_{\nu}G^{-1})(\partial_{\mu}G)]G^{-1}\,.
\end{eqnarray}
The commutator term transforms as
\begin{eqnarray} 
  i\,g_ag_b&&\!\!\!\!\!\![B'_{\mu},A'_{\nu}] = \\ \nonumber
  i\,g_ag_b&&\!\!\!\!\!\![(GB_{\mu}G^{-1}+(i/g_b)(\partial_{\mu}G)G^{-1}),
  \\ \nonumber &&\!\!\!\!(GA_{\nu}G^{-1}+(i/g_a)(\partial_{\nu}G)G^{-1})]
\end{eqnarray}
\begin{eqnarray} \label{commutator}
  &=& i\,g_ag_b G[B_{\mu},A_{\nu}]G^{-1} \\ \nonumber
  &-& G\{[G^{-1}(\partial_{\mu}G),g_aA_{\nu}] -
         [G^{-1}(\partial_{\nu}G),g_bB_{\mu}]\}G^{-1} \\ \nonumber
  &+& i\,G[(\partial_{\mu}G^{-1})(\partial_{\nu}G) -
           (\partial_{\nu}G^{-1})(\partial_{\mu}G)]G^{-1}\,.
\end{eqnarray}
Combining (\ref{dApdBp}) and  (\ref{commutator}) yields (\ref{Fintg}).

\end{document}